\documentclass{article} 
\usepackage{iclr2023_conference_tinypaper,times}

\usepackage{amsmath,amsfonts,bm}

\def\eqref#1{equation~\ref{#1}}

\def\1{\bm{1}}

\DeclareMathAlphabet{\mathsfit}{\encodingdefault}{\sfdefault}{m}{sl}
\SetMathAlphabet{\mathsfit}{bold}{\encodingdefault}{\sfdefault}{bx}{n}

\usepackage[hidelinks]{hyperref}
\usepackage{url}
\usepackage{multirow}
\usepackage{microtype}
\usepackage{arydshln}

\title{PhoWhisper: \\ Automatic Speech Recognition for Vietnamese}

\author{Thanh-Thien Le, Linh The Nguyen, Dat Quoc Nguyen \\
VinAI Research, Vietnam\\
  \small{\texttt{\{v.thienlt3, v.linhnt140, v.datnq9\}@vinai.io}}
}

\iclrfinalcopy 
\begin{document}

\maketitle

\begin{abstract}
We introduce \textbf{PhoWhisper} in five versions for Vietnamese automatic speech recognition. PhoWhisper's robustness is achieved through fine-tuning the Whisper model on an 844-hour dataset that encompasses diverse Vietnamese accents. Our experimental study demonstrates state-of-the-art performances of PhoWhisper on benchmark Vietnamese ASR datasets. We have open-sourced PhoWhisper at: \url{https://github.com/VinAIResearch/PhoWhisper}.
\end{abstract}

\section{Introduction}

Automatic speech recognition (ASR) technology, also referred to as speech-to-text, has experienced significant advancements \citep{wav2vec2-2020, seamlessm4t-2023, pratap2023scaling}, expanding its applicability across a wide range of applications. The state-of-the-art ASR model, Whisper \citep{whisper-2023}, has become extremely popular, being widely used in both academia and industry.

In this paper, we present an empirical study exploring Whisper for Vietnamese. Specifically, we further fine-tune the multilingual Whisper model on a large-scale ASR dataset that includes a diverse array of Vietnamese accents from different regions in Vietnam. This results in a fine-tuned model that we name PhoWhisper. Our empirical results demonstrate state-of-the-art performances of PhoWhisper, outperforming the previous best baselines on the Vietnamese Common Voice, VIVOS, VLSP 2020 Task-1 and VLSP 2020 Task-2 test sets.

We publicly release PhoWhisper, which can be used with \texttt{transformers}  \citep{wolf2019huggingface} and \texttt{openai-whisper} \citep{whisper-2023}.  We hope that PhoWhisper can serve as a strong baseline for future Vietnamese ASR  research and applications.




\section{PhoWhisper}
\label{sec:main}



Our PhoWhisper has five versions, including PhoWhisper\textsubscript{tiny}, PhoWhisper\textsubscript{base}, PhoWhisper\textsubscript{small}, PhoWhisper\textsubscript{medium} and PhoWhisper\textsubscript{large}, using the same
architectures of the multilingual models Whisper\textsubscript{tiny}, Whisper\textsubscript{base}, Whisper\textsubscript{small}, Whisper\textsubscript{medium} and Whisper\textsubscript{large-v2}, respectively.


\begin{table}[h]
\caption{Data statistics.}
\label{tab:data}
\centering 
\begin{tabular}{l|c|c|c|c}
\hline 
\multirow{2}{*}{\bf Dataset} & {\bf Training size}  & {\bf Validation size}  & {\bf Test size} & {\bf \#syllables in training set} \\ 
& (hours) & (hours) & (hours) & (min -- max $|$ average)\\
\hline 
CMV--Vi 14 & 3.04 & 0.41 & 1.35 & 1 -- 14 $|$ 7.55 \\
VIVOS  & 13.94 & 0.98 & 0.75 & 2 -- 30 $|$ 13.25 \\
\hdashline
VLSP 2020 Task-1 & 240.91 & 2.53 & 7.50 & \multirow{2}{*}{1 -- 349 $|$ 17.52} \\
VLSP 2020 Task-2 & -- & -- & 6.01 & \\
\hdashline
Our private data & 585.90 & -- & -- & 11 -- 24 $|$ 16.90 \\
\hline 
Total & 843.79 & 3.92 & 15.61 & --  \\
\hline 
\end{tabular}
\end{table}

We fine-tune our models on a large-scale ASR training set consisting of 844 hours of audio collected from four different resources, including \textbf{CMV--Vi}, the Vietnamese part of the Common Voice 14 \citep{commonvoice:2020}, \textbf{VIVOS} \citep{vivos2016luong}, VLSP 2020 ASR challenge,\footnote{\url{https://vlsp.org.vn/vlsp2020/eval/asr}} and our private data, as shown in Table \ref{tab:data}. Our ``private data'' is instrumental in providing the much-needed diversity of accents from 26K people spanning 63 provinces and municipalities, offering a profound understanding of the diverse ways in which Vietnamese is spoken. Finally, to enhance the robustness of our models against natural noises, we incorporate environmental sounds sourced from \citet{piczak2015dataset} and leverage the \texttt{audiomentations} library to add noise to half of the training set. That is, we randomly split the training set into two equal parts, A and B. We then augment part A with noise and combine the noise-augmented part A with the original part B to create the final training set of 844 hours of audio.

For fine-tuning, we use \texttt{transformers} \citep{wolf2019huggingface}, initializing PhoWhisper models from the corresponding multilingual Whisper models. We employ 8 A100 GPUs (40GB memory each) with a per-device batch size fixed at 4 and the number of gradient accumulation steps at 2 for all model versions, resulting in a global batch size of 64. The peak learning rates are set at 3.75e-5, 2.5e-5, 1.25e-5, 6.25e-6, and 5e-6 for  PhoWhisper\textsubscript{tiny}, PhoWhisper\textsubscript{base}, PhoWhisper\textsubscript{small}, PhoWhisper\textsubscript{medium}, and PhoWhisper\textsubscript{large}, respectively. We perform a total of 48,000 updating steps, which is approximately equivalent to 5 epochs.

\section{Empirical Results} \label{sec:exp}


We compare our models with the previous state-of-the-art ``wav2vec2''-based baselines from \citet{Thai_Binh_Nguyen_wav2vec2_vi_2021}. These baselines are obtained by first pre-training Wav2Vec2.0 ``base'' and ``large'' models \citep{wav2vec2-2020} on 13K hours of unlabeled Vietnamese YouTube audio and then fine-tuning them using 240+ hours of labeled training data from the VLSP 2020 ASR challenge.

\begin{table}[h]
\caption{Results on Vietnamese ASR benchmarks. ``\#paras'' denotes the number of parameters.}
\label{tab:main_results}
\centering 
\setlength{\tabcolsep}{0.4em}
\begin{tabular}{l|l|c|c|c|c}
\hline 
\multirow{2}{*}{\bf Model} & \multirow{2}{*}{\bf \#paras} & \multicolumn{4}{c}{\bf Word Error Rate} \\
\cline{3-6}
& & CMV--Vi & VIVOS & VLSP Task-1 & VLSP Task-2 \\
\hline
wav2vec2-base-vietnamese-250h & 95M& 102.04 & 10.83 & 21.02 & 50.35 \\
wav2vec2-base-vi-vlsp2020 & 95M & 103.71 & 9.90 & 16.82 & 44.91 \\
wav2vec2-large-vi-vlsp2020 & 317M & 101.41 & 8.61 & 15.18 & 36.75 \\
\hdashline
PhoWhisper\textsubscript{tiny} & 39M & 19.05 & 10.41 & 20.74 & 49.85 \\
PhoWhisper\textsubscript{base} & 74M & 16.19 & 8.46 & 19.70 & 43.01 \\
PhoWhisper\textsubscript{small} & 244M & 11.08 & 6.33 & 15.93 & 32.96 \\
PhoWhisper\textsubscript{medium} & 769M & \underline{8.27} & \underline{4.97} & \underline{14.12} & \underline{26.85} \\
PhoWhisper\textsubscript{large} & 1.55B & \textbf{8.14}  & \textbf{4.67} & \textbf{13.75} & \textbf{26.68} \\
\hline 
\end{tabular}
\end{table}

Table \ref{tab:main_results} presents our Word Error Rate (WER) results obtained for PhoWhisper and the baselines. Our PhoWhisper\textsubscript{small}, PhoWhisper\textsubscript{medium}, and PhoWhisper\textsubscript{large} outperform all the ``wav2vec2''-based baselines. Meanwhile, the remaining PhoWhisper\textsubscript{tiny} and PhoWhisper\textsubscript{base} are competitive with ``wav2vec2-base-vi-vlsp2020'' and perform better than ``wav2vec2-base-vietnamese-250h''. Here, PhoWhisper\textsubscript{large} establishes a new state-of-the-art WER score on each benchmark dataset.

\section{Conclusions}
In this paper, we have presented an empirical study exploring Whisper-based models, specifically PhoWhisper, for Vietnamese ASR. Our experimental results showcase PhoWhisper's state-of-the-art performance. We hope that our study and the public release of PhoWhisper will pave the way for further advancements and collaborations in this evolving field.


\subsubsection*{URM Statement}
The authors acknowledge that at least one key author of this work meets the URM criteria of ICLR 2024 Tiny Papers Track. 

\bibliography{phowhisper}
\bibliographystyle{iclr2023_conference_tinypaper}

%


\end{document}